\documentclass[sigconf]{acmart}

\usepackage[utf8]{inputenc}
\usepackage{graphics}
\usepackage{lipsum,booktabs}
\usepackage{CJKutf8}
\usepackage[T1]{fontenc}

\title{Query Suggestion for Click-Absent Queries in Enterprise Search}

\author{Gizem Gezici}
\author{Sabanci University, Istanbul, Turkey}
\author{Huawei Turkey R\&D Center, Istanbul, Turkey}

\begin{document}
\begin{CJK*}{UTF8}{gbsn}

\begin{abstract}
Creating alternative queries, also known as
query suggestion, has been proved to be helpful on improving
users' search experience. Owing to the suggestions, users could retrieve their information need more quickly and accurately. In many scenarios, these suggestions could be generated from the click-through logs by establishing a bipartite graph of the clicked query-document pairs. Most
of the existing methods focused on \emph{click-existing} queries which possess clicked information in the search logs, to suggest related queries using the co-clicked documents. In this paper, we propose a simple yet effective query suggestion method particularly for \emph{click-absent} queries by ensuring semantic consistency without utilising any additional resources. Our experimental results show that
the proposed technique generates comparatively good suggestions for \emph{click-absent} queries on a real bilingual enterprise search log.
\end{abstract}

\maketitle

\section{Introduction}
\label{sec:intro}
Search engines are ubiquitous. Currently, on average 3.5 billion Google searches are done per day~\cite{Internetlivestats}. Additionally, companies beneﬁt by in-house search platforms to easily access documents from various information sources. ``The application of information retrieval technology to information ﬁnding within organisations has become known as \emph{enterprise search}~\cite{EnterpriseSearch}''. It has been reported that in every 1.2 years, enterprise data worldwide is estimated to be doubled~\footnote{https://waterfordtechnologies.com/big-data-interesting-facts/} and employees spend 2.5 hours on average per day searching for information in a corporate environment~\cite{workerstime}.

The explosive growth of business data complicated the process of acquiring information in enterprise domain. When a user issues a query to a search platform, relevant documents from different sources are gathered, ranked, and displayed to the user for a given query. In IR, the performance of a search engine is evaluated by measuring its retrieval performance, i.e. retrieving the \emph{most relevant} documents for a user-speciﬁc query at higher positions in search results. The retrieval performance, i.e. satisfying the information need
of users, is measured via \emph{utility-based} evaluation measures, e.g. Precision at cut-off (P@$n$) and Discounted Cumulative Gain at cut-off (DCG@$n$) where $n$ is the number documents considered in the search results. However, it is crucial yet challenging for a user to generate a precise query that presents his/her information need.

The research on search logs found that user queries are usually short~\cite{wu2014improving}, 1.4 words are used on average~\cite{li2008characteristics}. The drawback is that
short queries often cannot provide sufﬁcient information to retrieve the corresponding relevant documents. For constructing queries more effectively, query suggestion
techniques have been utilised to recommend related queries~\cite{agrawal2011generating, craswell2012suggesting, wu2013suggesting}. The traditional query suggestion methods mainly ﬁnd candidate queries for a user query and score these queries to recommend the most relevant $k$ queries to the user. To calculate the similarity, researchers use the shared attributes of queries such as the number of times that two queries co-occur in a session, their number of common URLs, their text similarity etc. Although these methods have achieved good results for some queries, in most situations computing correlation scores of queries based on their characteristics cannot provide effective query suggestions since the queries are diverse, sparse and ambiguous~\cite{liu2017query}.

Query suggestion approaches could be mainly divided into two classes depending on the source of recommendation content, i.e. document-based and log-based methods~\cite{huang2003relevant}. The document-based methods extract related words to the query from the documents, whereas search logs are used for log-based query suggestion. However, document-based approaches suffer from rarely, i.e. \emph{long-tail}, or newly emerging queries, i.e. \emph{click-absent}. Moreover, search logs containing large amount of complete query data have been considered as the wisdow of crowds~\cite{fleenor2006wisdom} and could be mined to better reveal links between queries. Thus, the recent query suggestion methods mainly follow log-based approaches.

~\citet{mei2008query} employed random walk on query-URL bipartite graph to provide \emph{long-tail} suggestions, ordering candidate queries according to the walk time from the user query. The drawback of this approach is that \emph{long-tail} queries are less correlated with the
user queries. Thus,~\cite{liu2017query} recently presented a hybrid framework of random walk, topic concept extraction and suggestions from other four commercial search engines. Although the presented procedure can provide better suggestions for \emph{long-tail} queries and generate
suggestions even for \emph{click-absent} queries, this is only feasible when utilising the suggestions from other search engines. There are two downsides of this: i) The recommended suggestions are highly dependent on other search engines, and ii) The procedure brings additional overhead. For an enterprise search platform, suggesting queries as
highly depending on other engines may not provide domain-speciﬁc suggestions. Additionally, real search systems where user response time is also an issue, less complex suggestion procedure is more preferable.

In this work, we use the main idea in generating the preliminary candidate query set of~\cite{liu2017query} for \emph{click-existing} queries which uses random walk on a bipartite graph, without further enriching the set from topic concepts or mature search engines.

To alleviate the aforementioned drawbacks, we propose a new query suggestion method for \emph{long-tail} and \emph{click-absent} queries.

Our contributions can be summarised as follows:
\begin{itemize} \item We propose a simple yet effective query suggestion method particularly for \emph{click-absent} queries, which can also be used to enrich the candidate set for \emph{long-tail} ones, by exploiting the semantic information in the logs without using any additional resources.

\item We apply this framework on bilingual click logs to make a comparative evaluation of the state-of-the-art \emph{click-existing} and newly proposed \emph{click-absent} query suggestion procedure in a real enterprise search platform.
\end{itemize}

The remainder of the paper is structured as follows. In Section~\ref{sec:query_suggestion} we present the complete query suggestion framework and in Section~\ref{sec:query_suggestion_clickexisting} we detail the experimental setup. Then in Section~\ref{sec:conclusion_future}, we conclude the paper and give future work.

\section{Query Suggestion Framework}
\label{sec:query_suggestion}
In this chapter, we initially summarise the query suggestion procedure inspired by~\cite{liu2017query} for \emph{click-existing} queries and then elaborately mention our new approach for \emph{click-absent} queries.

\subsection{Query Suggestion for Click-Existing Queries}
\label{sec:query_suggestion_clickexisting}
The core approach is to construct a bipartite graph from the
\emph{clicked} query-document pairs in the click logs as done by~\cite{liu2017query}. We represent each query and document as a vertex and their click information, i.e. how many times this query is used to access this document (URL), as the corresponding edge weight in the graph as displayed in Fig. 1. Since this is a bipartite graph, there is no edge connecting two queries or two documents in the graph. In this way, we establish the graph using all the clicked query-document pairs in the log. After the graph construction, we compute similarity scores between queries based on the sample graph in Fig. 1 as follows. For instance, $q_1$ and $q_3$ are connected through URL $u_3$; the similarity between them is deﬁned as $W$($q_1$, $q_3$). $W$($q_1$, $q_3$) is computed as $e_{1,3}$*$e_{3,3}$ where $e_{1,3}$ is the edge weight between $q_1$ and $u_3$, and $e_{3,3}$ is
the weight between $q_3$ and $u_3$. Moreover, $q_1$ and $q_4$ are connected through URLs $u_2$ and $u_3$. Then the similarity between these queries, $W$($q_1$, $q_4$), is
computed as $e_{1,2}$*$e_{4,2}$ +
$e_{1,3}$*$e_{4,3}$. If there are more then two URLs in common, then the edge weights will be added to the sum. Lastly, $q_1$ and $q_5$ do not have any URL in common, i.e. queries are not connected. Thus, $q_1$ cannot be the suggestion query for the query  $q_5$ and vice-versa. Note that we use document titles instead of URLs for graph construction. Based on these, we computed the similarity scores between each pair of query. More rigorously, for a given query  $q_0$, we ﬁnd all of its connected queries in the graph as set $Q_{q_0}$ = {$q_1, q_2, q_3, ..., q_n$}. Then we sort these candidates by the similarity scores of $W$($q_1$, $q_n$) for the query $q_0$. In this work, we consider the top-10 connected queries as the ﬁnal candidates for the given query. Then, we store these query suggestions in our MongoDB database for each \emph{click-existing} query-document pair. Note that the technique inspired by~\cite{liu2017query} is applicable to the clicked pairs in the graph generating suggestions only for \emph{click-existing} queries. Yet, for a real search platform we need to ﬁnd query suggestions for \emph{click-absent} queries as well which we give details in the following part.

\begin{figure}[!t]
    \centering
    {\includegraphics{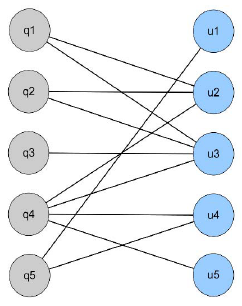}}
    \caption{Sample undirected Query-URL bipartite graph~\cite{liu2017query}.}
    \label{fig:dcgdiffstance}
\end{figure}

\subsection{Query Suggestion for Click-Absent Queries}
\label{sec:query_suggestion_clickabsent}
In this section, we present a new approach speciﬁcally for
\emph{click-absent} queries that do not have click information in the log. This is especially important in real applications since new queries are continuously emerging where users issue queries that are not in the click graph.

\begin{table*}[!t]
    \centering
    \caption{Query Suggestions for a Sample Set of~\emph{Click-Existing} Queries.}
    \begin{tabular}{c|c|c|c}
        \hline\hline
        Initial Query:~\emph{java} & Initial Query:~\emph{svn} & Initial Query:~\emph{java api} & Initial Query:~\emph{svn commit} \\
        \hline
        java 线程 (thread) & svn 使用(use) & java api 文档(documentation) & gerrit \\
        \hline
        \shortstack{spring \\ boot} & \shortstack{svn 安装指导 \\ (installation instructions)} & \shortstack{java api \\中文 (Chinese)} & \shortstack{svn get \\ lock commit} \\
        \hline
        \shortstack{最新的java技术 (latest \\
java technology)} & \shortstack{svn安装 \\ (installation)} & \shortstack{java 内存分析 \\(memory analysis)} & \shortstack{git commit \\
修改 (modification)} \\
        \hline
        \shortstack{c 语言 \\ (language)} & \shortstack{eclipse \\ svn} & \shortstack{java 多线程设计模式详解 \\
(multi-threaded design pattern)} & \shortstack{怎么删除git 分支 \\(how to
delete a git branch)} \\
        \hline
        java 8 & svn 服务器(server) & java excel & git review \\
        \hline\hline
    \end{tabular}
    \label{tab:existing}
\end{table*}

\begin{table*}[!t]
    \centering
    \caption{Query Suggestions for a Sample Set of~\emph{Click-Absent} Queries.}
    \begin{tabular}{c|c|c|c}
        \hline\hline
        \multicolumn{2}{c|}{\textbf{Initial Query: }$unix$} & \multicolumn{2}{c}{\textbf{Initial Query: }$java\: spring$} \\
        \hline
        \textbf{Similar Queries} & \textbf{Suggested Queries} & \textbf{Similar Queries} & \textbf{Suggested Queries} \\
        \hline
        dos2 unix & selenium python & java spring ioc & spring cloud \\\hline
        golang unix socket & jython 安装(installation) & powermock spring & aop \\\hline
        jython & python selenium & spring powermock & spring 注解 (annotations) \\\hline
        strcmp & dos2unix & spring aop & spring\\\hline
        cp 解压 (decompression) & linux rar 解压 (decompression) & spring 泛型 (generics) & spring echache \\
        \hline\hline
    \end{tabular}
    \label{tab:absent}
\end{table*}

Our method mainly combines the original approach mentioned in Section~\ref{sec:query_suggestion_clickexisting} and word embeddings to generate candidates for \emph{click-absent} queries. The main goal is to ﬁnd semantically similar queries of a given \emph{click-absent} query from the click graph. Then, we apply random walk on these similar queries in the already established bipartite
graph, instead of the given click-absent query to ﬁnd their connected queries. After that, we again sort the similarity scores of these connected queries and take the top-10 as the candidate query set for the given \emph{click-absent} query.

To ﬁnd semantically similar queries for a given query, we ﬁrstly train an ofﬂine Word2Vec CBOW model~\cite{mikolov2013efficient} on the search logs. Before the training, we apply some pre-processing steps on the corpus which will be mentioned in Section~\ref{sec:preprocessing} below. Then we obtain word embeddings and further generate centroid vectors of all queries using the embedding vectors suggested in~\cite{kuzi2016query}. After that, we ﬁnd similar queries in the graph for each \emph{click-absent} query of the click log. Based on the premise of generating a centroid vector as mentioned in~\cite{kuzi2016query}, we compute the centroid vector of a given query, \emph{java api} as follows. After having obtained the embeddings of the words \emph{java}
and \emph{api} separately with the Word2Vec model, we compute a mean vector on these word embeddings that is considered as the centroid vector for the query of \emph{java api}. We note that instead of the mean vector of the terms, one can also use the sum vector as done by~\cite{kuzi2016query}.

In the framework, semantically similar queries for all
\emph{click-absent} queries are found and stored in the database. Then using these similar queries, their connected queries in the graph are extracted with the aforementioned approach and also saved in the database.

\section{Experimental Setup}
\label{sec:experimental_setup}
In this section, we provide the description of our experimental setup. We ﬁrstly give details about the dataset and mention the pre-processing steps before applying the proposed query suggestion framework. Lastly, we make a comparative evaluation for the two query suggestion methods of \emph{click-existing} and \emph{click-absent} queries using human annotations.

\subsection{Dataset}
\label{sec:dataset}
The dataset is composed of the bilingual search logs from a one month period containing mixed English and Chinese queries/documents of an in-house enterprise search system in a company. The search system is similar to the Stack Overﬂow~\footnote{https://stackoverflow.com/} in which there are technical questions and their corresponding answers. In the platform, a user can search for an answer of a technical issue or write a new question.

Differently from a web commercial search engine, our dataset speciﬁcally holds two properties: i) Queries/documents come from a rather narrow domain, meaning that there are different query-document pairs coming from the related areas. ii) The majority of the search log instances contains highly domain-speciﬁc technical terms about the company's products and services. The ﬁrst property may help us on ﬁnding more
similar queries for the \emph{click-absent} part, whereas the second property may complicate the process of obtaining human labels with high quality for evaluation.

The dataset includes queries and document titles with click information, i.e. there is no document content. There are 328.362 query-document pairs in total. Out of this, 146.090 instances are unique non-clicked pairs where 115.682 of them are the click-through instances. There are 14.482 unique queries as 998 \emph{long-tail}, i.e. longer than 3 words
and clicked less than 5 times, whereas 286 of them are
\emph{click-absent}. 

\subsection{Pre-processing}
\label{sec:preprocessing}

We ﬁrstly found the unique query-document pairs by accumulating the click counts. Second, we removed punctuations and stop words in English and Chinese. Then, we made the queries and documents lower-cased and tokenised English with space/tab delimiters, whereas segmented Chinese content using the Stanford NLP Parser~\cite{manning2014stanford}. The parser works sufﬁciently well for Chinese segmentation with an accuracy of 93.65\%~\cite{tseng2005morphological}.

\subsection{Evaluation Results}
\label{sec:evaluation}
In this section, we evaluated the complete query suggestion framework. We initially established a bipartite click-graph from the pre-processed dataset, then for each \emph{click-existing} query we found their connected queries in the graph following the approach in Sect. II-A. For
the graph construction, we used only click-through logs of 115.682 unique query-document pairs. For \emph{click-absent} queries, we trained an ofﬂine Word2Vec CBOW model~\cite{mikolov2013efficient} to ﬁnd their semantically similar queries on the whole dataset of 146.090 instances after pre-processing. Then, for each \emph{click-absent} query we found the connected queries of its similar queries in the graph as proposed in Sect. II-B. In both suggestion methods, the candidate queries were sorted based on their
similarity scores and for each query the top-10 candidates were selected. For evaluation, the ﬁrst 5 candidates were chosen as displayed in Table~\ref{tab:existing} and Table~\ref{tab:absent}. In order to measure the performance, we used human annotations similarly to our reference paper~\cite{liu2017query}. Since our
essential aim is to improve user experience through query suggestion, in the scope of this work human annotations seem to be a proper evaluation method. We randomly selected 100 unique queries of \emph{click-existing} and 100 of \emph{click-absent} for evaluation. We then deﬁned scores 1
to 5 to evaluate the quality of the suggested queries. The score 5 shows that the initial query and the suggested query are the \emph{most} correlated, whereas score 1 means for the \emph{least} correlation. For each query, four of our colleagues scored the ﬁrst 5 suggestion results. Then, we computed an average value over these scores. In the scope of
this task, the annotators can be considered as experts since they possess computer engineering background and their level of English is sufﬁciently good. Nonetheless, this may still be a challenging task for the annotators where the dataset
contains English and Chinese words together as well as highly
domain-speciﬁc technical terms related to the products and services of the company. This may require additional knowledge apart from the general knowledge of engineering. Thus, to obtain labels with high quality we provided rules \& tips for the annotators as follows.

\begin{itemize}
\item You can give relevance scores of 1 to 5 as 1 (Not relevant) - 5
(Highly-relevant).

\item It would probably be better if you compare the suggested queries of
a given query among themselves.

\item You can use google translate for the Chinese content and google for
the domain-speciﬁc technical terms.

\item Please do not make any assumption about the suggested query list of
a given query; the order of the list is completely \emph{random}.

\item You can give the \emph{same} relevance scores for different suggested
queries of a given query.
\end{itemize}

Based on the average scores, we obtained the score of 3.98 for the \emph{click-existing} queries. This means that the expert annotators found 79\% of correlation between the initial query and its suggested queries extracted from the bipartite click graph. For the \emph{click-absent} queries, we got the score of 3.21 which corresponds to the correlation of 64\%. These results show that the query suggestion
method of \emph{click-existing} queries achieves a sufﬁciently good accuracy in a real enterprise search platform as well -- almost 80\% of correlation. Although the new approach shows a relatively weak correlation, it is over the correlation of 60\% which we believe to be sufﬁcient for the \emph{click-absent} queries. The relatively low score
is actually expected since we had to ﬁnd the connected queries of the semantically similar queries from the graph, instead of the given query itself due to the lack of click information. Note that our proposed method can be used for \emph{tail} queries, particularly for enriching the candidate set of the \emph{long-tail} queries which we leave as future work.

Lastly, we requested the annotators to give feedback related to the difﬁculty of the labelling task. They mainly raised two challenges as expected. First one was due to the mixed English and Chinese words in queries/documents. Second one was the existence of the domain-speciﬁc technical terms belonging to the company's products and services. Furthermore, annotators reported that the second labelling task, i.e. \emph{click-absent}, was more difﬁcult than the ﬁrst one. In this way, our evaluation results have been validated with the annotators' feedback.

\section{Conclusion \& Future Work}
\label{sec:conclusion_future}
In this work, we proposed a complete query suggestion framework for a real enterprise search system alleviating the drawbacks of the previous methods. We evaluated the state-of-the-art query suggestion method of \emph{click-existing} queries in an enterprise search platform and extended the method for \emph{click-absent} queries that are frequently encountered, especially in real search platforms.

We evaluated our framework on the real search dataset and make a comparative evaluation. The experiments showed that the suggestion method of \emph{click-existing} queries worked sufﬁciently good in a real bilingual enterprise search platform. Furthermore, this method yielded more correlated suggestions in comparison to the \emph{click-absent} one as expected since we could exploit only partial information -- there is no click data for the \emph{click-absent} queries as the name implies. 

In our experiments, we used Word2Vec for ﬁnding semantically similar queries and trained the model only on our corpus. Since the search log is composed of mixed English and Chinese words we could not use pre-trained models. Thus, as future work, we consider to use multilingual word embedding models that can be trained on different languages simultaneously. Lastly, we believe that our proposed approach can also be used to enrich the candidate query set for \emph{tail} queries which we left as future work.

\bibliographystyle{ACM-Reference-Format.bst}
\bibliography{main}

\clearpage\end{CJK*}
\end{document}